\documentclass[aps,prl,twocolumn,showpacs,superscriptaddress,groupedaddress,preprintnumbers,amsmath,amssymb]{revtex4-1}

\usepackage{graphicx}
\usepackage[amssymb]{SIunits}
\usepackage{ulem}
\usepackage{textcomp}
\usepackage{braket}
\usepackage{upgreek}

\usepackage[pdftex]{changebar}
\usepackage{xcolor}

\begin{document}

\title{Demonstration of a coherent electronic spin cluster in diamond}

\author{Helena S. Knowles}
\thanks{These authors contributed equally to this work.\\}
\affiliation{Cavendish Laboratory, University of Cambridge, JJ Thomson Ave, Cambridge CB3 0HE, UK}

\author{Dhiren M. Kara}
\thanks{These authors contributed equally to this work.\\}
\affiliation{Cavendish Laboratory, University of Cambridge, JJ Thomson Ave, Cambridge CB3 0HE, UK}

\author{Mete Atat\"{u}re}
\thanks{Electronic address: ma424@cam.ac.uk\\}
\affiliation{Cavendish Laboratory, University of Cambridge, JJ Thomson Ave, Cambridge CB3 0HE, UK}
\email{ma424@cam.ac.uk}

\date{\today}

\begin{abstract}
An obstacle for spin-based quantum sensors is magnetic noise due to proximal spins. However, a cluster of such spins can become an asset, if it can be controlled. Here, we polarize and readout a cluster of three nitrogen (N) electron spins coupled to a single nitrogen-vacancy (NV) spin in diamond. We further achieve sub-nm localization of the cluster spins. Finally, we demonstrate coherent spin exchange between the species by simultaneous dressing of the NV and the N states. These results establish the feasibility of environment-assisted sensing and quantum simulations with diamond spins.
\end{abstract}


\maketitle

NVs have emerged as a platform for magnetic field\cite{Taylor2008,Degen2008,Maze2008}, electric field\cite{Dolde2011} and temperature\cite{Neumann2013,Toyli2013} sensing, primarily because of their optical polarizability, long spin coherence times and stability in ambient conditions. Nanodiamonds (NDs) are versatile hosts for these sensors: They can be positioned to nanometre precision, integrated with a scanning probe\cite{Balasubramanian2008} and inserted into living cells\cite{McGuinness2011,Kucsko2013} with organelle specific targeting for in-situ probing of biological processes\cite{Krueger2012}. To achieve a workable yield of NVs a relatively high N concentration (10 - 500\,ppm) is required in the ND host. These dark N spins, each contributing an unpaired electron spin of S = $1/2$, produce a fluctuating magnetic field at the NV site that limits measurement sensitivity. However, by taking advantage of their coupling to the NV the dark spins can be turned into a resource, where the NV measurement signal is enhanced by contributions from these dark spins. An immediate opportunity is metrology with sensitivity scaling beyond the standard quantum limit\cite{Giovannetti2006,Goldstein2011,Cappellaro2012}. The implementation of such a scheme is challenging as it relies critically on the ability to generate long-lived polarization of a spin cluster that is coupled coherently to the NV spin. 

\begin{figure}[t!]
\centering
\includegraphics[width=0.48\textwidth]{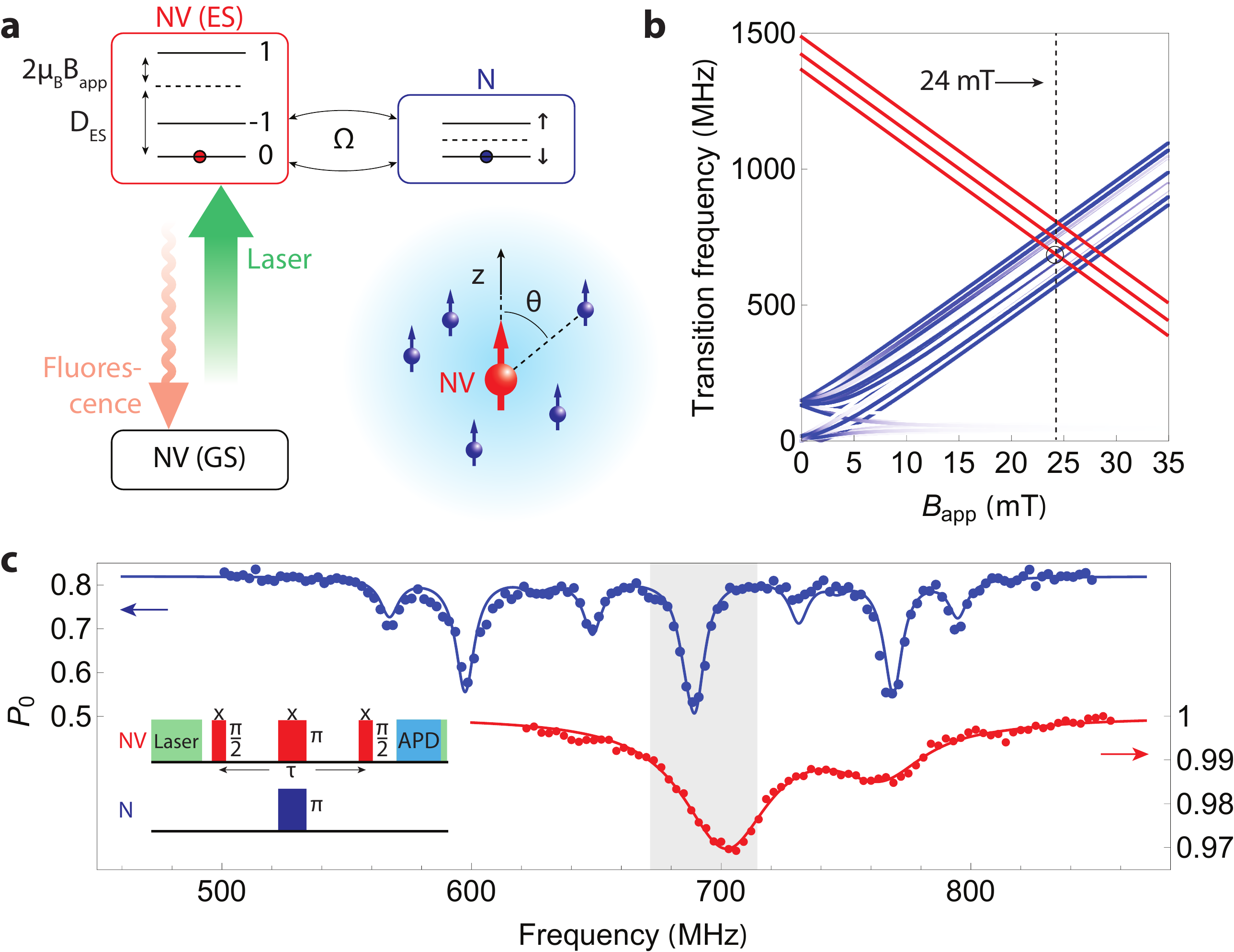}\\
\caption{(Color online) (a) Energy levels of the NV and N states. Spin transfer (at rate $\Omega$) between NV ES and N spin can take place efficiently, if the energy splittings are commensurate, Bottom right: illustration of the NV and proximal Ns. (b) The calculated NV ES electronic transitions (red) and N spin transitions (blue) as a function of applied field $B_{\rm{app}}$. (c) NV ES optically detected magnetic resonance data (red circles) and a triple Lorentzian fit (red curve). Spectrum of N spins (blue circles) measured through a DEER sequence illustrated in the inset: the NV GS spin echo (SE) coherence is monitored for fixed $\tau$ = 1\,$\upmu$s with NV spin rotations driven at 2.105\,GHz while a radiowave (RW) $\pi$-pulse (synchronous with the refocusing microwave (MW) $\pi$-pulse on the NV) is scanned in frequency and a model of driven N transitions (blue curve)\cite{Knowles2014}. At a magnetic field of 24\,mT the central transitions of the N spins ($m_{\rm{I}} = 0$) are commensurate with the lower, most highly populated NV hyperfine level as indicated by the grey region.}
\end{figure}

\begin{figure*}[t!]
\centering
\includegraphics[width=0.95\textwidth]{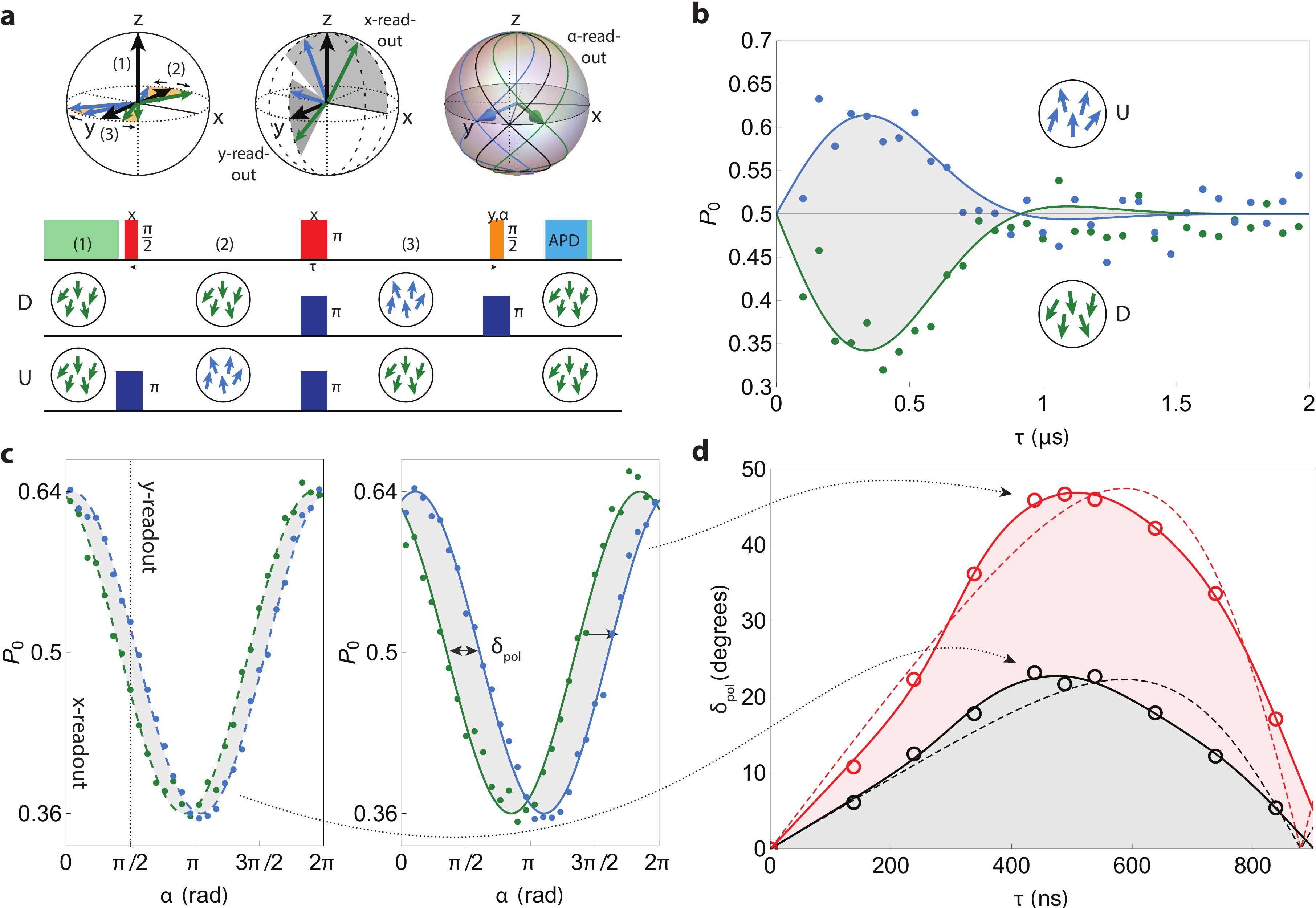}\\
\caption{(Color online) (a) Illustration of NV spin on the Bloch sphere as it undergoes a DSE measurement, along with the MW sequence used to control the NV spin (red, orange pulses) and the RW sequences for the N spins (blue pulses). The polarization of the N cluster (`down' green arrows and `up' blue arrows) occurs during the optical initialization step (1), the magnetic field created by the N spins is probed in (2) and (3) and the readout of the NV spin is performed after the final $\pi/2$ pulse. We perform two sequences labelled D and U where the NV detects the field created by a cluster polarized `down' (D) and `up' (U). The $\pi$ rotation on the N spin, coincident with the echo pulse, ensures that $B_{\rm{pol}}$ contributes a cumulative phase for the full DSE duration. The second pulse in D returns the cluster to its original state before the next measurement cycle. (b) DSE measurement with readout about the y axis, where we plot the probability of the NV state $\ket{0}$, $P_0$. The gap between data acquired for the sequence D (green) and U (blue) shows the presence of polarized spins surrounding the NV. Solid curves are fits to $\propto \frac{1}{2}\mathrm{exp}[-(\tau/t_\mathrm{DSE})^2] (1 + \mathrm{sin}\,\phi)$. (c) IDSE measurement where $\alpha$ is scanned with optical initialization time of 2\,$\upmu$s (left) and 20\,$\upmu$s (right). (d) Phase difference $\delta_{\rm{pol}}$ as a function of the IDSE free precession time $\tau$ for 2\,$\upmu$s (black) and 20\,$\upmu$s initialization (red). Fits using a three N spin model (solid lines, see main text for parameters) and a single N spin model (dashed lines).}
\end{figure*}

Due to their strong magnetic moment N impurities in nanodiamonds provide a naturally bound electronic spin ensemble, that is particularly attractive for magnetometry applications. Additionally, they allow the investigation of spin chains, proposed as information buses in quantum networks, and quantum simulators\cite{Plenio2007,Cai2013a}. In this Letter, we introduce a straightforward method that efficiently polarizes N impurities around a NV centre under ambient conditions. This is achieved by matching the NV optically excited state (ES) and the N spin transition energies with an applied magnetic field. Under continuous optical excitation the NV spin polarization is then transferred to the N cluster through dipolar interaction. We measure the resulting polarization by implementing an interferometric detection scheme on the NV ground state (GS). We are able to readout the number of individual spins within our cluster, their degrees of polarization and their coupling strengths to the NV. We also extract the location of the cluster spins to within a few lattice sites. Interleaving the ES polarization technique with the Hartmann-Hahn (HH) double resonance scheme allows us to address any proximal N spin in a full solid angle around the NV\cite{Hahn1950}. Finally, we use the HH scheme to reveal the coherent spin exchange dynamics between the NV and N spins.

\begin{figure}[t!]
\centering
\includegraphics[width=0.48\textwidth]{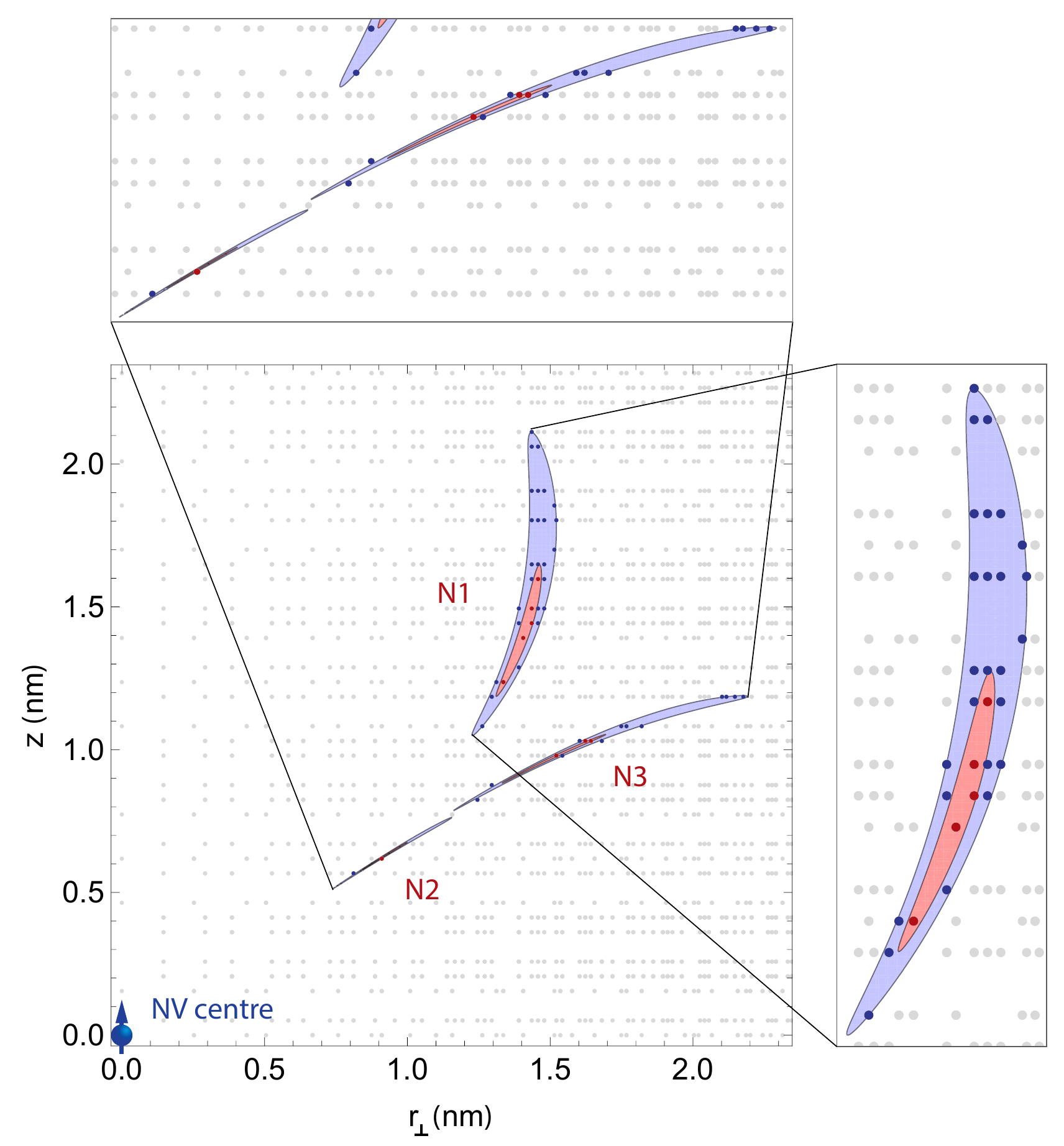}\\
\caption{(Color online) Map of N$_1$, N$_2$ and N$_3$ locations with respect to the NV centre. The N spins lie within the red areas (red circles) with 68\% probability and within the blue areas (blue circles) with 95\% probability. Grey circles represent the locations of carbon atoms in the diamond lattice. The axis $r_{\perp}=\sqrt{x^2+y^2}$ shows the transverse separation between the NV and the carbon atoms. In this representation each circle occurs 6 times within the 3-dimensional lattice, due to the diamond symmetry. }
\end{figure}

\it{Cluster polarization.}---\rm{We} work with diamond crystals of 20-nm mean diameter containing a single NV and $\sim$40~N impurities\cite{Knowles2014}. The NV  ($\rm{S}=1$ GS and ES, with \rm{I}~=~1 $^{14}$N nuclear host) is dipole coupled to a cluster of proximal N spins ($\rm{S}=1/2$, $\rm{I}=1$), as illustrated in Fig.~1(a). Dipole-dipole mediated spin exchange between the NV and N is suppressed due to an energy mismatch resulting from the NV zero-field splitting  $D_{\rm{ES}}\sim$ 1.4\,GHz. Hence, the NV spin polarization is not transferred to the dark spins. However, a magnetic field, $B_{\rm{app}}$, along the NV axis satisfying $D_{\rm{ES}}- 2(\mu_{\rm{NV}}/\it{h}) {B}_{\rm{app}} \sim \rm{2}(\mu_{\rm{N}}/\it{h})  {B}_{\rm{app}} $, ($\mu_{\rm{NV}}=\mu_{\rm{N}} = \mu_{\rm{B}}$) brings the two spins into resonance. We can treat the NV as a spin $1/2$ system with states $\ket{0}$ and $\ket{-1}$ ($\ket{+1}$ is far detuned) and the dipole interaction Hamiltonian reduces to
\begin{equation}
\mathcal{H}_{\rm{dip}} = \Omega (\rm{S_+^{N}}\rm{S_+^{NV}} + \rm{S_-^{N}}\rm{S_-^{NV}}) + \Delta\,\rm{S_z^N}\rm{S_z^{NV}}, 
\end{equation}
where $\rm{S_\pm}=(\rm{S_x}\pm i \rm{S_y})$, $\Omega = 3\, \mathrm{sin}^2\theta (\mu_0 \mu_\mathrm{B}^2)/(4 \pi r^3)$, $\Delta = (1- 3\, \mathrm{cos}^2\theta) (\mu_0 \mu_\mathrm{B}^2)/(4 \pi r^3)$\cite{Supplementary}. The state $\ket{0_{\rm{NV}},\uparrow_{\rm{N}}}$ evolves into $\ket{-1_{\rm{NV}},\downarrow_{\rm{N}}}$ at rate $\Omega$ enabling spin transfer between the two species. Figure~1(b) displays the calculated transition energies for $\ket{0} \leftrightarrow \ket{-1}$ in the NV ES manifold (red curves) and for N spin states $\ket{\uparrow} \leftrightarrow \ket{\downarrow}$ (blue curves) as a function of $B_{\rm{app}}$. Hyperfine interactions with the host $^{14}\rm{N}$ result in the three NV transitions and the signature N spectrum\cite{deLange2012,Knowles2014}. Figure~1(c) shows the NV ES optically detected magnetic resonance\cite{Fuchs2008} (red circles) acquired using continuous microwave and optical excitation at $B_{\rm{app}} =$ 24\,mT. The NV host nuclear spin shows polarization as expected at this $B_{\rm{app}}$\cite{Jacques2009,Smeltzer2009}. We measure the N spectrum (blue circles) using the double electron-electron resonance (DEER) scheme (Fig.~1(c) inset). The shaded area highlights the region of spectral overlap between the two species. 

\it{Cluster readout and localization.}---\rm{The} NV's repeated initialization into $\ket{0}$ combined with the resonant interaction pumps dark spins preferentially into the $\ket{\downarrow}$ state. To measure polarization build-up we probe the magnetic field along the NV axis, $B_{\rm{pol}}$, generated by the N electron cluster. 
This enables the direct measurement of the NV environment polarization rather than inferring it indirectly from loss of NV coherence during spin exchange\cite{Laraoui2013,Belthangady2013}.
We use a directional spin echo sequence (DSE) to reveal the sign of the field (in contrast to the spin echo used in the DEER scheme which only reveals its magnitude). This is achieved by performing the last $\pi/2$ rotation about the y axis, as illustrated in the first two Bloch spheres in Fig.~2(a). The DSE signal is given by $S_{\rm{DSE}} \approx\frac{1}{2}\mathrm{exp}[-(\tau/t_\mathrm{DSE})^2] (1 + \mathrm{sin}\,\phi)$, where $t_\mathrm{DSE}$ is the NV dephasing time, $\phi =  \mu_{\rm{B}} B_{\rm{pol}} \tau/{\hbar}$ is the phase acquired by the NV during $\tau$ and $S_{\rm{DSE}} =1$ (0) corresponds to NV state $\ket{0} (\ket{1})$. The top pulse train in Fig.~2(a) shows the NV sequence and the next one (D) shows the cluster spin rotations (driven at 689 MHz) with the resulting polarization orientations. Figure~2(b) shows the DSE contrast signal (blue circles) as a function of the free precession time, $\tau$, following 2\,$\upmu$s of optical initialization. Pulse trace U in Fig.~2(a) is a complimentary sequence where the N spins are flipped before the DSE measurement starts. This polarization inversion yields a phase with opposite sign (green circles in Fig.~2(b)), confirming that the $B_{\rm{pol}}$ we detect is only due to polarized spins.

To extract $\phi$ in a manner robust to NV spin dephasing ($t_\mathrm{DSE}$) and pulse errors we next introduce a new technique, interferometric DSE (IDSE): Fixing the time delay, $\tau$, and scanning the phase, $\alpha$, of the last $\pi/2$ pulse (see fig.~2(a)) produces interference fringes, whose phase is given by $\phi$. Figure~2(c) shows IDSE after 2\,$\upmu$s (left panel) and 20\,$\upmu$s of optical polarization (right panel) with $\tau$ = 400 ns. The phase difference, $\delta_{\rm{pol}} = 2 \mu_B B_{\rm{pol}}\tau/\hbar$, between the D (blue circles) and U (green circles) sequences increases from 23 to 46 degrees, corresponding to -2.5\,$\upmu$T and -5\,$\upmu$T net field along the NV axis. This indicates that cluster polarization builds up on $\upmu$s timescales. 

We can now probe the time evolution of the NV spin in its polarized environment to reveal the couplings and degrees of polarization of individual spins in the cluster. The black (red) open circles in Fig.~2(d) show $\delta_{\rm{pol}}$ as a function of $\tau$ for 2-$\upmu$s (20-$\upmu$s) optical excitation. We model this evolution with a system of $n$ N spins interacting with the NV\cite{Supplementary}. The maximum phase value is linked to the degree of polarization in the $\ket{\downarrow}$ state, $p_{\rm{i}} = $ $ \rm{P}_{\ket{\downarrow}} - \rm{P}_{\ket{\uparrow}}$, of each N spin N$_{\rm{i}}$, where $p_{\rm{i}}$ is $100$\% for $\ket{\rm{s}}_{\rm{N}}$ in state $\ket{\downarrow}$  and 0\% for $\ket{\rm{s}}_{\rm{N}}$ unpolarized. The rise and fall time is determined by $\Delta_{\rm{i}}$ of each N$_{\rm{i}}$ to the NV. 
This coupling strength corresponds to a magnetic field $B_{\rm{i}}$ of $-\frac{ \it{h} \rm{\Delta_{\rm{i}}}}{\upmu_{\rm{N}}}$ for the state $\ket{\downarrow}$ and $\frac{\it{h} \rm{\Delta_{\rm{i}}} } {\upmu_{\rm{N}}}$  for  $\ket{\uparrow}$ at the site of the NV.

To illustrate the strong dependence of the phase evolution on the number of N spins we show a best fit achieved for a single dark spin. The fit is the dashed black (red) curve in Fig.~2(d) for 2-$\upmu$s (20-$\upmu$s) polarization, which shows an  $n=1$-model does not explain our data. To determine the number of N spins in our system, the goodness of fit with increasing $n$ must be balanced against over-parameterisation. This problem is addressed using the Akaike information criterion and we find that our data is best described by an $n=3$ model with over 98$\%$ likelihood, yielding  $\Delta_1$ = -1.64$\pm$0.09\,MHz, $\Delta_2$ = 870$\pm$80\,kHz, $\Delta_3$ = 530$\pm$30\,kHz, $p_1$~=~2 $\pm$ 1\,\%, $p_2$~=~7 $\pm$ 1 \,\%, $p_3$~=~31 $\pm$ 1\,\% (2-$\upmu$s data, solid black curve) and $p_1$'~=~5 $\pm$ 1\,\%, $p_2$'~=~7 $\pm$ 2\,\%, $p_3$'~=~72 $\pm$ 2\,\% (20-$\upmu$s data, solid red curve)\cite{Supplementary}. The distinct angle dependence of $\Delta$ and $\Omega$ (linked to polarization) provides a direct link to the location of the spins. We note that N spins with $\Omega \lesssim 1/T_{\rm{1,N}} \sim 100\,\rm{kHz}$ and $\Delta \lesssim 1/T_{\rm{DSE}}\sim\rm\,{1.5\,MHz}$ do not contribute to the IDSE signal, where $T_{\rm{1,N}}$ is the typical N spin lifetime. As $\Delta$ and $\Omega$ both scale as $1/r^3$, distant N spins can be neglected.

The ability to locate dark spins is crucial when designing and optimizing a quantum protocol. In order to determine the locations of the spins from our measurement, we use a time-dependent model of an optically pumped NV coupled to a N spin. This allows the extraction of $\Omega_{\rm{i}}$ from the rise in polarization $p_{\rm{i}}$ of each spin N$_{\rm{i}}$ as a function of optical polarization time\cite{Supplementary}. Combining this with $\Delta_{\rm{i}}$ we are able to single out a few lattice sites as the probable locations for spins N$_1$, N$_2$ and N$_3$. The remarkable precision reached is displayed in Figure 3, where the red (blue) dots indicate atomic sites within the 68\% (95\%) probability bound. More statistics would improve the spatial resolution reported here.

\it{Coherent exchange between NV and N electron spins.}---\rm{Environment-assisted} sensing schemes exploit the collective coherence of the initially polarized NV-cluster to achieve enhanced sensitivity\cite{Goldstein2011}. We probe this coherence with the Hartmann-Hahn (HH) double locking sequence: The NV GS and N spins are driven into dressed states with identical transition energies in the locking frame, thereby allowing spin exchange via $\Delta$\cite{Hahn1950}. Figure~4(a) shows an illustration of the pulse sequences used for the HH coupling experiment. The top sequence shows the locking pulses for the N spin and the lower three illustrate the two directional sequences (D$_+$ and D$_-$) and the alternating sequence (A) performed on the NV. By comparing the D$_+$ and D$_-$ signals we probe the degree and direction of polarization of the N cluster, whereas sequence A allows probing of the spin cluster, irrespective of its polarization.

\begin{figure}[t!]
\centering
\includegraphics[width=0.43\textwidth]{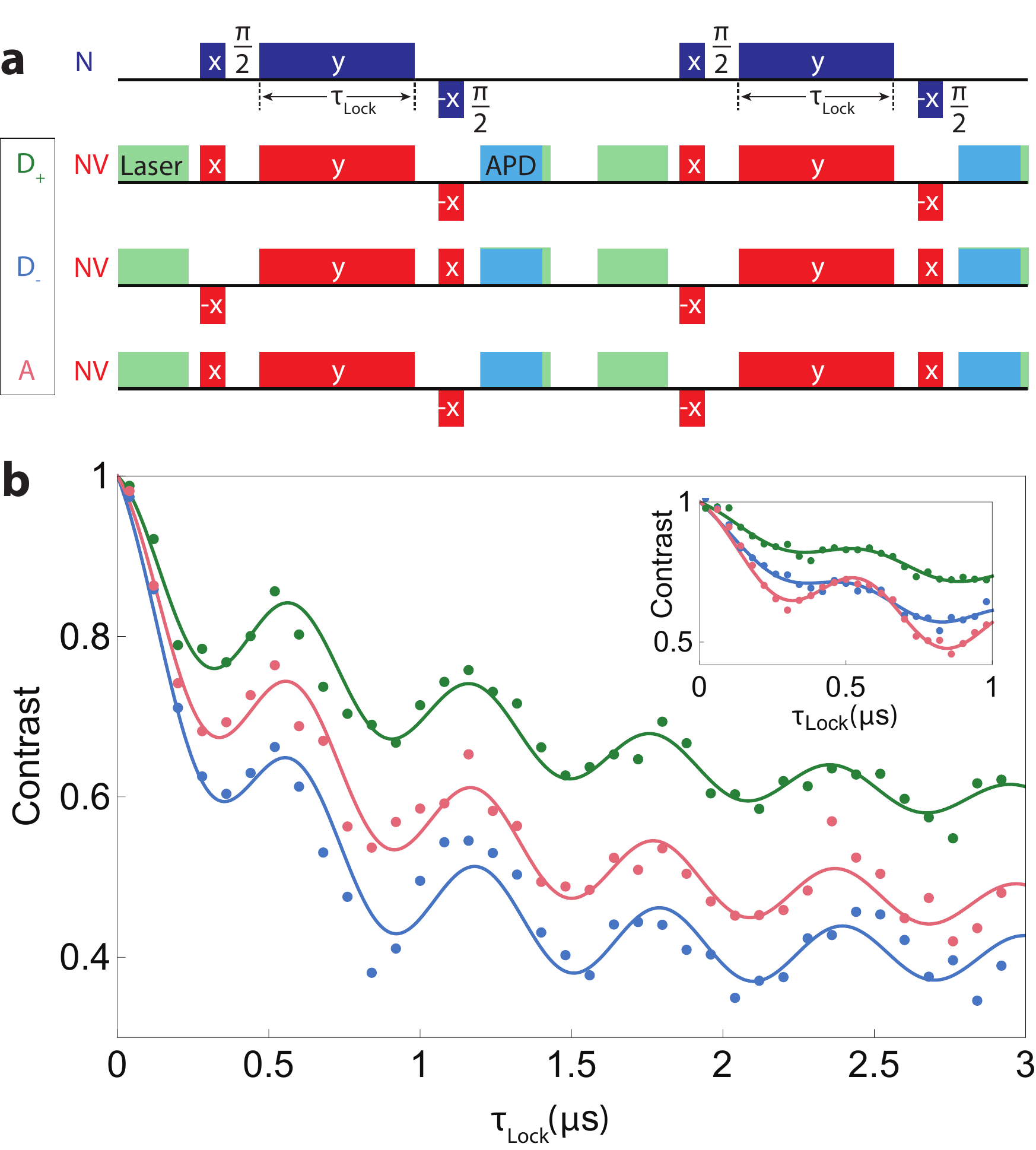}\\
\caption{(Color online) (a) Illustration of sequences used for the double locking HH experiments. Top sequence: N spin is locked about the y axis for duration $\tau_{\rm{Lock}}$ after the first $\pi_{\rm{x}}/2$ rotation and then returned to its initial position with a $-\pi_{\rm{x}}/2$ pulse. Three sequences below: i) The NV spin is locked into the $\ket{+} = \frac{\ket{0}+\rm{i}\ket{1}}{\sqrt{2}}$ state (D$_+$), ii) into the $\ket{-}= \frac{\ket{0}-\rm{i}\ket{1}}{\sqrt{2}}$ state (D$_-$) and iii) the lock is alternated between the $\ket{+}$ state and the $\ket{-}$ state (A). 
 (b) NV contrast as a function of $\tau_{\rm{Lock}}$ with predominantly optical polarization of the N cluster (20\,$\upmu$s optical initialization) for D$_+$ (green circles), D$_-$ (blue circles) and A (red circles). All signals are normalized to the NV lock decay in the absence of sequence N. The solid curves are fits to $a\,\rm{Cos}[2\pi\nu\tau_{\rm{Lock}}] e^{-\tau_{\rm{Lock}}/t_{\rm{osc}}} + (1-\it{a}-C) e^{-\tau_{\rm{Lock}}/t_{\rm{lock}}} + C$ yielding oscillation amplitudes of $a =$ 0.085, 0.094 and 0.095 for D$_+$, A and D$_-$, respectively. This indicates that N$_1$ is coupled to the NV with 1.68 $\pm$ 0.01\,MHz and is only weakly polarized by the optical scheme. Inset: Spin locking experiment for mixed optical and locking frame HH polarization (2\,$\upmu$s optical initialization), where N$_1$ is now polarized by the HH locking (amplitudes for D$_+$ and D$_-$, and A are $a =$ 0.033, 0.038 and 0.088, respectively). In the D$_+$ sequence, the optical and HH polarization act in the same direction and we observe a stronger polarization than in b), while in D$_-$ they compete with each other and the spin transfer is suppressed compared to b).}
\end{figure}

Figure~4(b) displays the HH contrast acquired with sequences D$_+$, D$_-$ and A for 20 $\upmu$s optical excitation. In the directional sequences, the optically polarized spins either block (D$_+$, green circles) or allow (D$_-$, blue circles) interactions with the NV, leading to different observed decay levels. Crucially, we see an oscillation corresponding to spin exchange between the NV and a N spin within the cluster with a coherence time of 2.3\,$\upmu$s. This reveals the presence of a coherent combined state and the extracted oscillation frequency of 1.68 $\pm$ 0.01\,MHz is consistent with $\Delta_1$, obtained independently from IDSE. The coherent oscillation is independent of the applied locking sequence, indicating weak optical polarization of N$_1$. This is in agreement with our findings from IDSE.

The directional HH sequence can be used to polarize spins with finite $\Delta$~\cite{Hahn1950,London2013}. Combining the HH sequence with our optical scheme we can polarize cluster spins through both $\Delta$ and $\Omega$ couplings and address spins irrespective of their orientation to the NV. In the inset of Fig.~4(b) we excite for 2\,$\upmu$s only, to allow both HH and optical polarizations to contribute. The HH sequence now polarizes N$_1$ to $p_1$ = 65 $\pm$ 13\,\%, as manifested by the reduced oscillation amplitudes in D$_+$ and D$_-$. By changing the ratio between optical polarization 
and HH polarization the cluster can thus be polarized in a spatially targeted fashion enabling selective spin pumping (see Ref. \cite{Supplementary} for details).

\it{Conclusions.}---\rm{The} degree of polarization reached for spin N$_1$ corresponds to an external magnetic field of over 300~T under ambient conditions or to a temperature of 20~mK at our $B_{\rm{app}} = 24$\,mT.
This ability to prepare and our IDSE readout technique could enable the use of dark spins as spatially distributed ancillary sensors for high resolution imaging of single molecules.
The robust coherent exchange observed between two spins could provide a platform for quantum enhanced sensing\cite{Goldstein2011,Cappellaro2012}. The operation of this system under ambient conditions and the modest requirements on magnetic field magnitude and alignment precision also make it suitable for sensing in living biological samples. The engineering of large N clusters localized around NVs in diamond would provide a system with more ancillary spins and thus stronger enhancement in environment-assisted schemes. Further, coupled electron spins in diamond combined with the control demonstrated in this Letter offer an opportunity to investigate quantum state transfer in spin chains experimentally. In the longer term, such spin systems, with bright spin nodes coupled by chains of dark spins could enable the realization of quantum simulators on chip at room temperarure\cite{Bose2003,Yao2012,Ajoy2013}.

\begin{acknowledgments}
We gratefully acknowledge financial support by the Leverhulme Trust Research Project Grant 2013-337 and the Winton Programme for the Physics of Sustainability. H.S.K. also acknowledges financial support by St John's College through a Research Fellowship. We are grateful to A. Nunnenkamp and R. Stockill for fruitful discussions. 
\end{acknowledgments}


%

\end{document}